\documentclass[a4paper,11pt]{article}
\usepackage{pos}

\usepackage{physics}
\usepackage{mathtools}

\usepackage{booktabs}
\usepackage{array}
\usepackage{adjustbox}
\usepackage{cleveref}

\newcolumntype{C}[1]{>{\centering\arraybackslash}m{#1}}
\newcolumntype{L}[1]{>{\arraybackslash}m{#1}}

\newcommand{\ii}{\,\mathrm{i}\,}

\title{Status of double virtual NNLO QCD corrections for high multiplicity processes}

\author*[a,b]{Vasily Sotnikov}

\affiliation[a]{Physik-Institut, University of Zurich,\\
  Winterthurerstrasse  190, Zurich, Switzerland}

\affiliation[b]{Department of Physics and Astronomy, Michigan State University,\\
567 Wilson Road, East Lansing, USA}

\emailAdd{vasily.sotnikov@physik.uzh.ch}

\abstract{
Pushing the reach of NNLO QCD predictions to $2\to3$ production processes is one of
the pillars of precision phenomenology program at the LHC. In this talk we will
overview recent results and developments in the calculation of two-loop five-point
amplitudes contributing towards achieving this goal. We will discuss challenges
encountered in advancing the state-of-the-art beyond the class of massless
five-point scattering. We will then present a basis set of transcendental
functions sufficient to express any planar two-loop five-particle scattering
amplitude with one external massive leg. This basis greatly facilitates
derivation of compact analytic form of scattering amplitudes, and opens a
possibility of their fast and reliable numerical evaluation. Applications for
phenomenology of electroweak boson production can be reasonably anticipated in
the near future.
}

\FullConference{%
  Loops and Legs in Quantum Field Theory - LL2022,\\
  25-30 April, 2022\\
  Ettal, Germany
}

\begin{document}

\renewcommand{\hookAfterAbstract}{%
  \par\bigskip
  ZU-TH 35/22, MSUHEP-22-026
}
\renewcommand{\logo}{\relax}

\maketitle

\section{Introduction}
\label{sec:intro}

Precise theoretical predictions for $2\to3$ scattering cross sections are key ingredients for achieving the ultimate precision goal at the Large Hadron Collider.
The computation of higher order terms of their series expansion in the Standard Model coupling constants is therefore essential.
To reach accuracy of theoretical predictions at the level of few percent at least NNLO QCD and NLO EW corrections are commonly required.
In this report we will focus on the former.

To calculate a differential hard-scattering cross section to higher orders in perturbation theory one needs to compute all relevant loop scattering amplitudes,
and a framework to arrange the intricate cancellation of their infrared divergences. It has been recently demonstrated that the latter is possible 
for the $2\to3$ processes with the most complicated IR structure \cite{Czakon:2021mjy,Chen:2022ktf} at NNLO QCD.
On the other hand, only the simplest two-loop amplitudes required for NNLO QCD corrections to five-point scattering are currently known (see \cref{tab:soa}).
Therefore the availability of NNLO predictions for high-multiplicity processes is largely limited by our ability to calculate two-loop amplitudes.

\section{State of the art}

Scattering amplitudes are calculated in perturbation theory as sums of a large number of Feynman integrals.
The latter can be reduced to a smaller set of \emph{master} integrals by systematically resolving integral relations such as integration-by-parts (IBP) identities.
This procedure is highly challenging for problems with many kinematic scales involved, mainly due to intermediate expression swell.
Thanks to the remarkable progress in the reduction of multiscale amplitudes, 
most of the \emph{massless} two-loop five-point amplitudes relevant for hadron collider phenomenology has been computed in the past two years in leading-color approximation (see \cref{tab:soa} for references). 
First results in full QCD have also been reported \cite{Agarwal:2021vdh,Badger:2021imn}. 
Among others, one of the main advances which made these analytic calculations possible was 
systematic integration of exact numerical evaluations over finite fields and functional reconstruction techniques \cite{vonManteuffel:2014ixa,Peraro:2016wsq,Peraro:2019svx} into 
modern amplitude reduction frameworks \cite{Peraro:2019svx,Abreu:2020xvt}.

\begin{table}[ht]
  \centering
  \begin{tabular}{L{16ex}C{16ex}*{3}{C{12ex}}}
    \toprule
      & Comment   & Complete analytic results & Public numerical code & Cross sections \\
    \midrule
    $pp\to jjj$ & l.c. & \cite{Abreu:2021oya} & \cite{Abreu:2021oya}  &  \cite{Czakon:2021mjy,Chen:2022ktf} \\
    $pp\to \gamma\gamma j$ & l.c.$^\star$  & \cite{Agarwal:2021grm,Chawdhry:2021mkw} & \cite{Agarwal:2021grm} & \cite{Chawdhry:2021hkp} \\
    $pp\to \gamma\gamma\gamma$ & l.c.$^\star$ & \cite{Abreu:2020cwb,Chawdhry:2020for} & \cite{Abreu:2020cwb} & \cite{Chawdhry:2019bji,Kallweit:2020gcp} \\
    $pp\to \gamma\gamma j$ &  & \cite{Agarwal:2021vdh}  &  & \\
    $gg \to  \gamma\gamma g$ & NLO loop induced  & \cite{Badger:2021imn}  & \cite{Badger:2021imn}  & \cite{Badger:2021ohm} \\
    \midrule
    $pp \to W b \bar{b}$ & l.c.$^\star$, on-shell $W$ & \cite{Badger:2021nhg} &  & \\
    $pp \to W(l\nu) b \bar{b}$ & l.c. & \cite{Abreu:2021asb,Hartanto:2022qhh} &  & \cite{Hartanto:2022qhh} \\
    $pp \to W(l\nu) j j$ & l.c. & \cite{Abreu:2021asb} &  & \\
    $pp \to Z(l \bar{l}) j j$ & l.c.$^\star$ & \cite{Abreu:2021asb} &  & \\
    $pp \to W(l\nu) \gamma j$ & l.c.$^\star$ & \cite{Badger:2022ncb} &  & \\
    $pp \to H b \bar{b}$ & l.c., $b$-quark Yukawa  & \cite{Badger:2021ega} &  & \\
    \bottomrule
  \end{tabular}
  \caption{
    Known two-loop QCD corrections for five-point scattering processes at hardon colliders.
    ``l.c.''\ refers to the calculations in the leading-color approximation; ``l.c.$^\star$'' means that in addition non-planar l.c.\ contributions are omitted.
    All public codes employ \texttt{PentagonFunctions++}  \cite{Chicherin:2020oor,Chicherin:2021dyp} for numerical evaluation of special functions.
  }
  \label{tab:soa}
\end{table}

The full potential of the finite-field reconstruction methods can be achieved if the final analytic expressions for an amplitude simplify dramatically compared to intermediate expressions.
Mathematical understanding of relevant Feynman integrals and the associated space of transcendental functions is essential to expose these simplifications.
The knowledge of all \emph{pure} two-loop five-point massless master integrals \cite{Abreu:2018aqd,Chicherin:2018old}, and
the basis of relevant transcendental functions, \emph{pentagon functions} \cite{Chicherin:2020oor}, was indeed a prerequisite for the success of analytic multiscale amplitude calculations.
For phenomenological applications of loop scattering amplitudes, it is crucial that their numerical evaluation is sufficiently fast and stable, such that Monte Carlo integration over final-state phase space is feasible.
Satisfying this requirement is highly nontrivial, especially for the transcendental parts of amplitudes.
Indeed, for two-loop five-point scattering, applications of the methods commonly employed in organizing the function spaces of lower-multiplicity amplitudes has not been effective to date.
Instead, a new method of function-basis construction, based on properties of \emph{iterated integrals}, and its numerical evaluation was developed \cite{Chicherin:2020oor} to open the door for broad phenomenological applications.

Going beyond purely massless two-loop five-point amplitudes, first analytic results with one external mass (we will refer to it as \emph{one-mass} kinematics here)
have started to appear \cite{Badger:2021nhg,Abreu:2021asb,Hartanto:2022qhh,Badger:2022ncb,Badger:2021ega}.
A steep increase in complexity has been observed. In particular, the number of numerical samples that is required to reconstruct the analytic form of rational coefficients 
using the ``black-box'' functional reconstruction algorithms becomes prohibitively large.
However the general methods developed for tackling the massles five-point amplitudes could be still applied after certain improvements 
of analytic reconstruction techniques \cite{Badger:2021nhg,Abreu:2021asb}.
With new possibilities of systematically constraining the analytic form of rational coefficients being currently explored \cite{DeLaurentis:2020qle,DeLaurentis:2022otd},
one can be cautiously optimistic that even more complex process can be tackled within this framework in the future.

As in the fully massless case, the theoretical understanding of two-loop five-point one-mass Feynman integrals has been essential.
Canonical differential equations for all planar two-loop five-point one-mass integral topologies \cite{Abreu:2020jxa}, as well for some nonplanar topologies \cite{Abreu:2021smk} are already known,
however a complete set of nonplanar internal topologies is still under investigation.
Representations of all integrals with planar topologies and certain nonplanar integrals through Goncharov polylogarithms (GPLs) have been found \cite{Papadopoulos:2015jft,Canko:2020ylt,Kardos:2022tpo}.
Nevertheless, finding adequate representations in physical regions still remains challenging.
A special function basis for strictly color-ordered amplitudes have been constructed in \cite{Badger:2021nhg}, and its numerical evaluation through generalized series expansions \cite{Hidding:2020ytt} has been explored.
A complete planar function basis was constructed in \cite{Chicherin:2021dyp}, and it was demonstrated that remarkably fast and stable numerical evaluation of these functions is possible. 
We discuss this in \cref{sec:pentagon-functions}.
These developments have been applied to achieve the first complete calculation of NNLO QCD corrections involving one-mass kinematics, $W b \bar{b}$ production at the LHC \cite{Hartanto:2022qhh}.
It is worth noting that important advances in methods of numerical evaluation of Feynman integrals have been also achieved.
In particular, a direct numerical evaluation of the two-loop five-point one-mass integrals is now feasible \cite{Liu:2021wks,Hidding:2022ycg}.
These methods provide valuable input for future analytic studies of even more complex integrals.

\newcommand{\CA}{\mathcal{A}}

\section{Reduction of multiscale loop amplitudes}
\label{sec:rational}

To compute a loop scattering amplitude, one starts by writing down its integrand which schematically can be represented as
\begin{equation}\label{eq:integrand}
  {\CA}(\ell_l)=\sum_{\Gamma\in\Delta}
  \sum_{i}c_{\Gamma,i} (\vb*{s}, \epsilon) \; 
  \frac{m_{\Gamma,i}(\ell_l)}{\prod_{j\in P_\Gamma}\rho_j}\,,
\end{equation}
where $\ell_l$ denotes the loop momenta of the problem, $\Delta$ is the set of distinct propagator structures $\Gamma$,
and $P_\Gamma$ is the multiset of inverse propagators $\rho_j$ in $\Gamma$.
$m_{\Gamma,i}(\ell_l)$ are polynomials in loop momenta and rational functions of external kinematics $\vb*{s}$ and the dimensional regulator $\epsilon = \frac{1}{2} (4-D)$.
Integral reduction brings the amplitude's integrand to the form 
\begin{equation}\label{eq:A}
  \CA = \sum_{i}c_{i}(\vb*{s},\epsilon)~\mathcal{I}_{i}\,,
\end{equation}
where $\mathcal{I}_{\Gamma,i}$ are the master integrals and $c_{i}(\vb*{s},\epsilon) \in \mathbb{Q}(\vb*{s},\epsilon)$.
The master integrals are then expanded in $\epsilon$ and expressed through special functions.
If these functions form a basis, i.e.\ they are \emph{algebraically} independent,
the UV and IR divergences can be subtracted analytically and one can derive the finite remainder,
\begin{equation}
  \mathcal{R} = \sum_{\vb*{i}} r_{\vb{i}}(\vb*{s}) ~ \vb*{g}^{\vb*{i}} + \order{\epsilon},
\end{equation}
where $\vb*{g}^{\vb*{i}}$ in our case are the relevant monomials of pentagon functions in the multi-index notation.
For a well-chosen basis $\vb*{g}^{\vb*{i}}$ the rational coefficients $r_{\vb{i}}(\vb*{s})$ are significantly simpler than any intermediate expressions encountered
in their derivation starting from the amplitude's integrand in \cref{eq:integrand}.

An impressively powerful method to take advantage of the finite remainders' simplicity is to reconstruct their analytic expressions from
exact numerical samples over finite fields \cite{vonManteuffel:2014ixa,Peraro:2016wsq}.
The complete amplitude reduction frameworks built around this idea were made publicly available:
\texttt{FiniteFlow} \cite{Peraro:2019svx,Badger:2017jhb,Badger:2019djh} and two-loop numerical unitary \cite{Ita:2015tya,Abreu:2017xsl,Abreu:2017idw,Abreu:2017hqn,Abreu:2018jgq,Abreu:2020xvt}.
In the latter, even the integrand in \cref{eq:integrand} is evaluated numerically, eliminating the need for analytic processing of individual Feynman diagrams.

Employing the algorithm of \cite{Peraro:2016wsq}, an analytic reconstruction of the rational functions $r_{\vb{i}}(\vb*{s})$ with no a priori knowledge of their structure is possible.  
This is known as black-box reconstruction. 
An important advantage of the black-box reconstruction algorithm of ref.\ \cite{Peraro:2016wsq} is that the computational complexity of the algorithm
itself is negligible compared to the complexity of obtaining numerical samples of $r_{\vb{i}}(\vb*{s})$.
On the other hand, the number of required samples is maximal.
Planar two-loop five-point massless amplitudes can be obtained with a straightforward application of some variations of the black-box reconstruction.
More complicated amplitudes, however, would require too many numerical samples.
For example, a black-box reconstruction of the two-loop amplitudes contributing to $pp\to Wjj$ production would require $10^7$ samples \cite{Abreu:2021asb}, with each sample taking an order of few minutes.
It is therefore clear that black-box reconstruction cannot be straightforwardly applied beyond purely massless five-point scattering.

Fortunately, the coefficients $r_{\vb{i}}(\vb*{s})$ in scattering amplitudes are not arbitrary rational functions. On the contrary, their analytic
structure is deeply constrained by the underlying physics. The finite-field framework opens up a possibility to constructively incorporate constraints 
on their analytic structure and therefore reduce the number of numerical samples required for their reconstruction.
For instance, the denominators of the coefficients $r_{\vb{i}}$ are entirely determined by the rational subset of the associated symbol alphabet \cite{Abreu:2017hqn}.
Another example is that reconstruction and partial fractioning in one of the variables can allow one to fix the coefficients of a dense polynomial ansatz in the remaining variables with a significantly smaller number of required samples \cite{Badger:2021imn}. 
In addition, the tractable size of the polynomial ansatz can be significantly increased by the Vandermonde sampling \cite{Abreu:2021asb}, 
which allows to invert a linear system of size $N$ in $\mathcal{O}(N^2)$ time, instead of $\mathcal{O}(N^3)$.
Armed with these ideas, the number of samples required to reconstruct the two-loop amplitudes contributing to $pp\to Wjj$ production could be decreased by two orders of magnitude.
More systematic approaches in constraining the ansätze from their behavior in special limits are also being explored \cite{DeLaurentis:2020qle,DeLaurentis:2022otd}.
It would be interesting to see if they can be applied in the calculations of yet unknown amplitudes in the future.

Finally, let us make an important observation that the form of coefficients $r_{\vb{i}}$ that we obtain from any of the modern reconstruction 
algorithms is by far not ideal. In fact, it has been observed \cite{Abreu:2019odu} that dramatic simplification can be achieved 
with \emph{multivariate} partial fractioning \cite{Abreu:2019odu,Boehm:2020ijp,Heller:2021qkz,Bendle:2019csk}.
These simplifications are crucial to achieve sufficient numerical performance for phenomenological applications.  
Due to the unpredictable complexity of Gröbner bases calculation, this step itself might become challenging in applications beyond purely massless five-point scattering.
In the meantime, one might wonder if it is possible to find a way to reconstruct the simplified form of the result directly?
This intriguing possibility would significantly advance the capabilities of current computational methods.

Other key developments in IBP reduction techniques concern judicious selection and organization of the IBP identities \cite{Gluza:2010ws,Ita:2015tya,Larsen:2015ped,Georgoudis:2016wff,Boehm:2018fpv,Agarwal:2020dye,Guan:2019bcx}.
A particularly helpful observation in this context is to systematically avoid introducing integrals with propagators raised to higher powers.
The latter are ubiquitously encountered in generic IBP identities, but only sparsely appear in integrands of scattering amplitudes.
Eliminating the unnecessary integrals from the identities beforehand can significantly reduce the complexity of the residual linear systems.
These ideas have been instrumental in many of the recent two-loop five-point amplitude calculations.

\section{Planar one-mass pentagon functions}
\label{sec:pentagon-functions}

As we argued in the previous sections, knowledge of a transcendental function basis is advantageous both for studying the analytic structure of scattering amplitudes and for their efficient numerical evaluation.
In this section we present our method of constructing such a basis for five-point scattering with one external mass at two loops. We will refer to this basis as \emph{one-mass pentagon functions}.
Our method is largely based on the earlier work on the massless pentagon functions in ref.\ \cite{Chicherin:2020oor}, and described in detail in ref.\ \cite{Chicherin:2021dyp}.
Here we will mostly focus on highlighting the differences and new developments in the latter compared to the former.

Master integrals can be systematically expressed in terms of \emph{pure} functions order-by-order in $\epsilon$ within the method of canonical differential equations.
The canonical DEs for integrals $\vec{f}_{\tau,\sigma}$ of the integral topology (or familty) $\tau$ and permutation of external momenta $\sigma$ take the form
\begin{align}\label{eq:cDE}
  \dd \vec f_{\tau,\sigma}  = \epsilon \, \dd A_{\tau,\sigma} \vec f_{\tau,\sigma}, \qquad \dd  A_{\tau,\sigma}  = \sum_i a^{(i)}_{\tau,\sigma} \, \dd\log W_i \,,
\end{align}
where $a^{(i)}_{\tau,\sigma}$ are rational matrices, and $W_i$ are letters from the symbol alphabet.
For amplitudes with few scales an exceptionally successful strategy for construction of function bases has been to map the $\epsilon$-expansion of Feynman integrals into a particular class
of functions known as \emph{multiple polylogarithms} (MPLs). These functions can be viewed as iterated integrals over logarithmic one-forms with particular structure,
\begin{equation} \label{eq:mpl-kernels} 
  \dd{\log(t - w(\vb*{s}))} = \frac{\dd{t}}{t - w(\vb*{s})},
\end{equation}
where the pole positions $w(\vb*{s})$ may depend on the scattering kinematics. 
Within the DE approach, one can attempt to find a path $\gamma : [0,1] \to \mathcal{P}$, where $\mathcal{P}$ is the space of kinematic variables,  such that the pullbacks of all $\dd{\log}$ forms in the r.h.s.\ of \cref{eq:cDE} 
take the form of \cref{eq:mpl-kernels}. Practical considerations aside, this is always possible for rational alphabets. In multiscale problems, however, algebraic letters are ubiquitous.
Maximal cuts of relevant integrals typically evaluate to a square root $r$, and letters of the form $\frac{q - r}{q + r}$ with $q$ polynomial get introduced.
It is then clear that finding coordinate charts and/or appropriate paths on which all logarithmic forms from the alphabet are of the form of \cref{eq:mpl-kernels} is highly nontrivial.\footnote{%
  The question of whether such mapping for all Feynman integrals exists is, in fact, an open problem.
}
There has been much effort dedicated to addressing this problem \cite{Heller:2019gkq,Heller:2021gun,Bonetti:2020hqh,Kreer:2021sdt,Duhr:2021fhk,Papadopoulos:2015jft,Canko:2020ylt}.
Nevertheless, it becomes increasingly clear that even if one succeeds in deriving some MPL-representation of master integrals, finding a good representation in the whole physical region is very challenging (see e.g.\ 
discussion in \cite{Papadopoulos:2015jft,Canko:2020ylt, Gehrmann:2018yef, Canko:2020ylt, Duhr:2021fhk,Chaubey:2022hlr}).  
Typical issues are proliferation of spurious branch cuts, which makes analytic continuation difficult, and explosion of number of MPL functions required to represent the solution.
This leads to unacceptably slow evaluation in the physical region, even in the simplest five-point massless case.
In addition, perhaps somewhat counter-intuitively, this has an effect that MPL representations can, in fact, obscure analytic properties of amplitudes, and constructing a basis may be not straightforward.

The considerations above motivate us to follow a different route and forgo the attempts of mapping Feynman integrals into MPLs, even in cases when it may be in principle possible.
Instead, we can trivially solve the \cref{eq:cDE} through Chen's iterated integrals \cite{Chen:1977}, and construct a basis of needed functions relying only on the properties of iterated integrals.
We proceed as follows.
First, we write the solutions of \cref{eq:cDE} \cite{Abreu:2020jxa} from a single point $X_0$ in the physical scattering region for all permutations of external momenta $\sigma$ (such that we cover all possible scattering amplitudes) and for all relevant integral topologies $\tau$,
\begin{align}
\vec f^{(w)}(X) = \sum_{w' = 0}^{w} \; \sum_{i_1,\ldots,i_{w'} = 1}^{108} \vec \kappa^{(w-w')}_{i_1,\ldots,i_{w'}} \left.[W_{i_1},\ldots,W_{i_{w'}}]\right._{X_0}\!\!(X) \,. \label{f_it_int_gen}
\end{align}
where $\kappa^{(w-w')}$ are transcendental constants of weight $w-w'$,
\begin{align}
\kappa^{(w-w')}_{i_1,\ldots,i_{w'}} = a^{(i_1)}a^{(i_2)} \ldots a^{(i_{w'})} \vec{f}^{(w-w')}(X_0) \,, \label{kappa}
\end{align}
and the iterated integrals of weight $w$ along the path $\gamma$ are defined recursively as
\begin{align}
\left.[W_{i_1},\ldots,W_{i_{w}}]\right._{X_0}\!\!(X) = \int_\gamma \dd\log W_{i_w}(X') \, \left.[W_{i_1},\ldots,W_{i_{w-1}}]\right._{X_0}\!\!(X'), \quad []_{X_0} = 1 \,. \label{it_int_def}
\end{align}
To obtain the values $\vec{f}^{(w)}(X_0)$ of master integrals at the initial point $X_0$, we employ their representations through GPLs from ref.\ \cite{Canko:2020ylt}, and identify algebraic relations with PSLQ algorithm.
The representation \eqref{f_it_int_gen} allows us to define a transcendental-weigh-graded vector space 
\begin{equation}
  \mathbf{G}  = \bigoplus_w \mathbf{G}^{(w)} \coloneqq  \bigoplus_w \mathrm{span}\{\vec f^{(w)}\}.
\end{equation}
Linear independence of iterated integrals with different sequences of logarithmic kernels allows us to straightforwardly construct a basis in $\mathbf{G}^{(w)}$.
In addition, the shuffle product of iterated integrals which maps $\mathbf{G}^{(w_1)}\otimes \mathbf{G}^{(w_2)} \to \mathbf{G}^{(w_1+w_2)}$ allows us to recursively construct an \emph{irreducible} basis,
i.e.\ the basis which is linearly independent from any products of lower-weight functions. In other words, the irreducible basis functions are algebraically independent.

In practice, it is convenient to first construct a basis under the symbol map, which simply projects out all but the longest iterated integrals at each transcendental weight. 
One can then assign beyond-the-symbol terms to the basis functions in a way such that the basis dimension does not increase, and the basis functions are still related to the (monomials of) $\epsilon$-expansion
of master integrals. This has two important advantages. First, the expressions of master integrals through basis functions do not involve unnatural constants as was the case in \cite{Chicherin:2020oor}, and
only $\zeta$-values and $\ii \pi$ are explicitly present, as expected.\footnote{%
  As a consequence, we also do not have to introduce constants which are odd under changes of square-root signs.
}
Second, the action of the permutation group of external momenta on the one-mass pentagon functions is well-defined and can be straightforwardly derived. This makes it possible to cross scattering amplitudes into 
different channels by simple substitutions, instead of performing a rather intricate procedure as described in \cite{Abreu:2021oya}.
The results of our construction of one-mass pentagon functions are summarized in \cref{tab:classificationcounting}.

\begin{table}[ht]
  \centering
  \begin{tabular}{c*{3}{C{13ex}}}
    \toprule
    Weight & Linearly independent & Irreducible & Cyclic\\
    \midrule
    1 & 11 & 11 & 6 \\
    2 & 86 & 25 & 8\\
    3 & 483 & 145 & 31\\
    4 & 1187 & 675 & 113\\
    \bottomrule
  \end{tabular}
  \caption{
    Number of master integral components at each $\epsilon$ order. The second column corresponds to the number of the $\mathbb{Q}$-linearly independent components.
    The third column shows the number of irreducible components, i.e.\ the number of one-mass pentagon functions. 
    The third column corresponds to the number of one-mass pentagon functions contributing to strictly color-ordered amplitudes.
    The total number of topologically-independent master integrals  is 1417.
}
\label{tab:classificationcounting}
\end{table}

There is a great degree of ambiguity in choosing the basis elements. Compared to \cite{Chicherin:2020oor}, we take advantage of this ambiguity to make certain properties of scattering amplitudes explicit. 
For example our weight 4 basis has the following structure:
\begin{equation}
\resizebox{0.95\hsize}{!}{$
\overbrace{1,\ldots,67,\underbrace{68,\ldots,106}_{{\cal S}}, \underbrace{107,\ldots,112}_{\mathcal{Z}},\underbrace{113}_{{\cal S}, \sqrt{\Delta_5}}}^{\text{cyclic}},\overbrace{114,\ldots,441,\underbrace{442,\ldots,664}_{{\cal S}}, \underbrace{665,\ldots,672}_{\sqrt{\Delta_5}}, \underbrace{673,\ldots,675}_{{\cal S},\sqrt{\Delta_5}} }^{\text{non-cyclic}}
$} \,.
\end{equation}
Here functions with $\sqrt{\Delta_5}$ involve the letter which equals the pentagon gram determinant. These functions are expected to drop out of properly defined finite remainders (see the discussion in \cite{Chicherin:2020umh}).
The functions involving the subset of letters denoted by $\mathcal{Z}$ were observed to not contribute to any of the known two-loop amplitudes \cite{Badger:2021nhg}, an interesting fact which that has no known explanation. 
We also maximally isolate the functions containing letters which can vanish in the physical region (denoted as the set $\mathcal{S}$) to improve numerical performance.
The cyclic subset of functions was already studied in \cite{Badger:2021nhg} and is sufficient for any strictly color-ordered amplitude.

\begin{figure}[ht]
  \centering
  \includegraphics[width=0.3\textwidth]{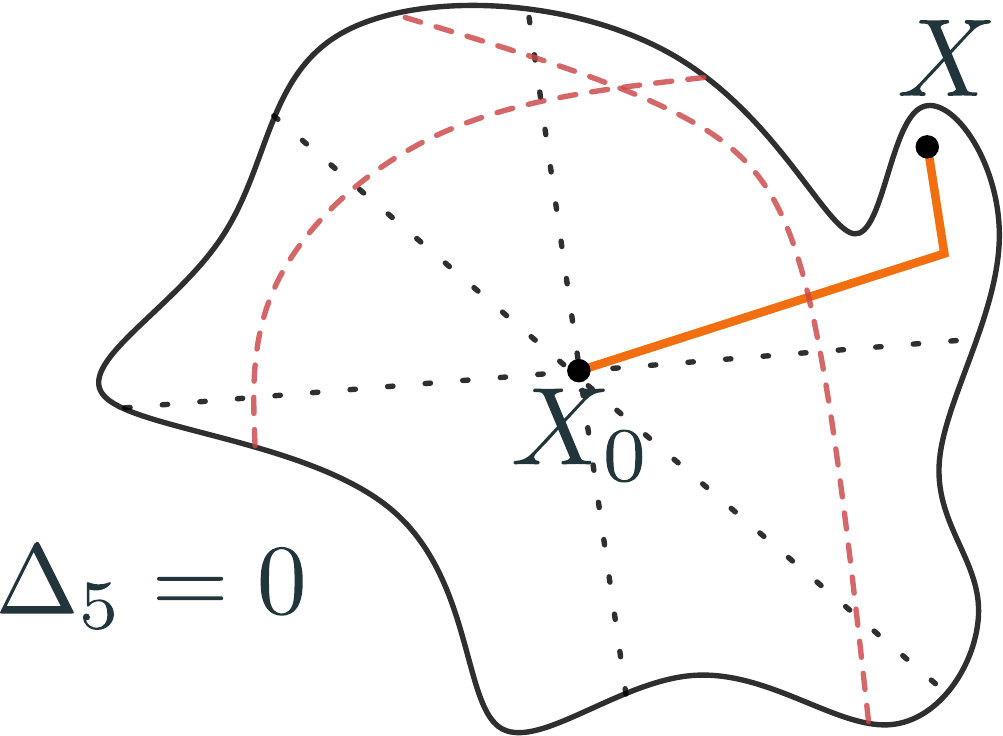}
  \caption{A sketch of the numerical integration over a path connecting the points $X_0,X ~\in \mathcal{P}$, bounded by the equation $\Delta_5=0$.
    The initial point $X_0$ is chosen to lie on the intersection of zero-sets of all linear letters that can vanish within the physical region, denoted by dotted black lines.
    The dashed red lines correspond to quadratic letters that can vanish within the physical region.
  }
  \label{fig:path-cartoon}
\end{figure}

Our strategy of numerical evaluation of one-mass pentagon largely follows the one from \cite{Chicherin:2020oor} and is inspired by \cite{Caron-Huot:2014lda,Gehrmann:2018yef}.
We derive their explicit representations in terms of logarithms and dilogarithms up to weight two, making sure that all the functions are well-defined in the whole physical region.
For weight three and four we derive one-fold integral representations, and we explicitly check that the integrands are well-defined analytic functions in the whole integration domain.\footnote{%
  Logarithmic integrable singularities can occur at the endpoints of the integration point, which does not create additional complications due to our choice of the numerical integration algorithm. 
}
We evaluate the integrals by a specifically optimized numerical quadrature algorithm which guarantees exponential convergence.
We encounter two new complications compared to ref.\ \cite{Chicherin:2020oor} which are schematically illustrated in \cref{fig:path-cartoon}.
First, it is possible that a point $X$ cannot be connected by a straight line that lies entirely within the physical phase space $\mathcal{P}$.
To avoid this situation we chose an additional random point $X^\prime \in \mathcal{P}$, such that the line segments $[X_0,X^\prime]$ and $[X^\prime, X]$ both lie within $\mathcal{P}$.
Second, in the presence of \emph{quadratic} letters that can vanish within $\mathcal{P}$, we must take care of spurious singularities. To this end,
we find series expansions of the integrands in the neighbourhood of vanishing quadratic letters, thereby achieving analytic cancellation of spurious poles.

\begin{figure}[ht]
  \centering
  \includegraphics[width=0.8\textwidth]{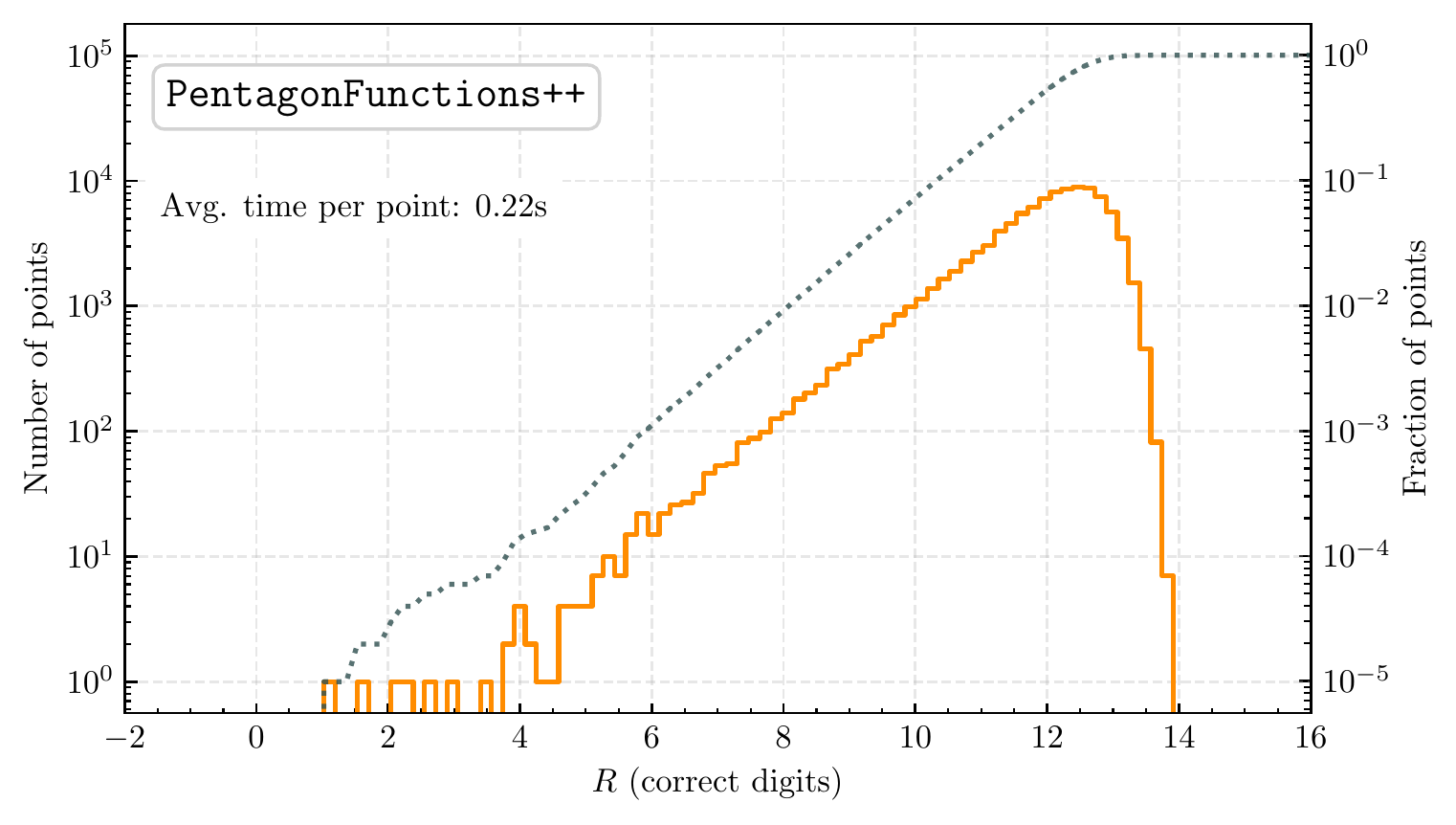}
  \caption{
    Logarithmic distribution of minimal correct digits of one-mass pentagon functions evaluated on a sample of $10^5$ phase-space points.
    The average evaluation time of all pentagon function in double precision on a single thread is estimated on a server with \textit{Intel(R) Xeon(R) Silver 4216 CPU @ 2.10GHz}.
    This figure is taken from ref.~\cite{Chicherin:2021dyp} licensed under \textit{CC-BY 4.0}.
  }
  \label{fig:num_stab}
\end{figure}

We demonstrate the excellent numerical performance of one-mass pentagon functions by evaluating them on $10^5$ points sampled from a typical physical phase-space 
(see the details in ref.\ \cite{Chicherin:2021dyp}).
We estimate the relative error $r_i$ of a double-precision evaluation of a pentagon function by comparing it to its evaluation in quadruple precision.
We then define the number of correct digits as $-\log_{10}\abs{r_i}$.
On each phase-space point $X$ we evaluate all one-mass pentagon functions and we take the worst (smallest) number of digits $R$ among them as the overall measure of accuracy for this point.
We display the distribution of $R(X)$  in \cref{fig:num_stab}.
We observe excellent numerical stability in the bulk of the phase space and average evaluation time of less than a second. 
Clearly, one-mass pentagon functions completely eliminate the problem of numerical evaluation of special functions relevant for two-loop five-point one-mass scattering for any future phenomenological applications.

\section{Conclusions}

The availability of NNLO QCD predictions for $2\to 3$ processes still remains limited by our ability to compute two-loop five-point amplitudes.
Much progress has been achieved in the past few years. 
Calculations of leading color amplitudes for the production of up to three jets and photons have been completed, and the first results in full color have started to appear.
Beyond purely massless $2\to 3$ scattering a steep increase in complexity is observed, but the techniques developed for the massless calculations still remain promising.  
The most complicated kinematics currently within reach is five-point with one external mass, and the analytic results for leading-color $Vjj$, $Wjy$, $Hb\bar{b}$ amplitudes have been derived.
The planar function basis for this class of processes has been successfully constructed, enabling phenomenological applications in the near future. 

Despite the great progress, NNLO QCD corrections for many key $2\to 3$ Standard Model processes ($Vjj,~Z b \bar{b},~Hjj,~t\bar{t}j,~ t \bar{t}V,~t \bar{t} H$, etc.) remain unknown, 
and it is very much expected that new challenges will need to be overcome to complete calculations of all processes relevant for the LHC physics program.
Nevertheless, the modern computations methods do not appear to be entirely stretched to their limits. 
We can therefore remain carefully optimistic that the ultimate precision reach of the LHC will not be limited by the uncertainties due to truncation of the perturbation series.

\paragraph{Funding information}

This work has received funding from the European Research Council
(ERC) under the European Union’s Horizon 2020 research and innovation programme grant
agreement 101019620 (ERC Advanced Grant TOPUP).

\bibliographystyle{JHEP}
\bibliography{main.bib}

\end{document}